\documentclass[nofootinbib]{revtex4}

\usepackage{graphicx}
\usepackage{dcolumn}
\usepackage{bm}
\usepackage{color}
\usepackage{bbold}

\begin{document}

\markboth{ A. Sol\'e, X. Oriols,  D. Marian and N. Zangh\`i}{How does Quantum Uncertainty Emerge from Deterministic Bohmian Mechanics?}

\title{How does Quantum Uncertainty Emerge from Deterministic Bohmian Mechanics?}

\author{A. Sol\'e}

\address{Departament de Filosofia, Universitat de Barcelona, Spain.}

\author{X. Oriols}

\address{Departament d\rq{}Enginyeria Electr\`{o}nica, Universitat Aut\`{o}noma de Barcelona, Spain.}

\author{D. Marian}

\address{Dipartimento di Ingegneria dell'Informazione, Universit\`a di Pisa. Via G. Caruso 16, 56122 Pisa, Italy.}

\author{N. Zangh\`i}

\address{Dipartimento di Fisica dell'Universit\`a di Genova and INFN sezione di Genova, Italy.}

\begin{abstract}
Bohmian mechanics is a theory that provides a consistent explanation of quantum phenomena  in terms of  point particles whose motion is  guided by the wave function. In this theory, the state of a system of particles is defined by the actual positions of the particles and the wave function of the system; and the state of the system evolves deterministically.  Thus, the Bohmian state  can be compared with  the state in classical mechanics, which is given by the positions and momenta of all the particles, and which also evolves deterministically.  However, while in classical mechanics it is usually taken for granted and considered unproblematic that the state is, at least in principle, measurable, this is not the case in Bohmian mechanics. Due to the linearity of the quantum dynamical laws, one essential component of the Bohmian state, the wave function, is not directly measurable. Moreover, it turns out that the  measurement of the other component of the state---the positions of the particles---must be mediated by the wave function; a fact  that in turn implies that the positions of the particles, though measurable,  are constrained by {\em absolute uncertainty}. This is the key to understanding how Bohmian mechanics, despite being deterministic, can account for all quantum predictions, including quantum randomness and uncertainty.
\end{abstract}

\maketitle

\section{Introduction}
\label{sec1}

Most quantum experiments produce results that cannot be predicted with certainty.  No matter how we improve our experimental techniques, this uncertainty will not diminish.  Any attempt to explain why this is so will of course depend on which theory is used to describe the quantum world. In this paper, we will explain the origin of quantum randomness and uncertainty from the standpoint of adopting (non-relativistic) Bohmian mechanics\cite{bohm52,durr13,Oriols12} as that theory.

Both classical mechanics and Bohmian mechanics are deterministic theories. This means that the specification of the complete state of a (classical or Bohmian) system at one time \textit{$t_0$}, together with the laws of the theory, is compatible with only one state of the system at any other time $t \neq t_0$. So, if the complete state of a (classical or Bohmian) system is specified at one time, the whole history of the system is determined. Obviously, the two theories differ in terms of their laws and also in what needs to be specified in order to provide a complete characterization of a system. In classical mechanics, the state of a system is fully characterized if the velocities and positions of the particles that constitute the system are provided at an initial time. In contrast, a complete characterization of a Bohmian system requires both the initial positions of its constituent particles to be given and also the initial wave function that guides them.

Very often, determinism is associated with absolute predictability. However, this association is unjustified. For instance, if a classical system is chaotic, since its state can only be known with a finite degree of precision, its evolution after a period of time will be uncertain. Moreover, if the system is macroscopic and has some $10^{23}$ particles, measurement of all the initial conditions (the position and the velocity of each particle in the system) is not practically possible. But, without knowledge of the initial conditions that completely characterize the state of the system, its future evolution cannot be predicted. These sources of unpredictability in classical systems are also found in Bohmian mechanics. There are, however, two fundamental differences between classical and Bohmian unpredictability. 

The first difference has to do with the experimental accessibility of the state, i.e., the initial conditions. We usually determine the initial conditions of a system by measuring them. In classical mechanics---and despite all the practical difficulties mentioned above---the idea that the initial conditions of a system can (theoretically) be measured is not problematic. We will show that in Bohmian mechanics this is not the case. It follows from the dynamical laws of Bohmian theory themselves that the wave function of a system cannot be measured. In addition, if the wave function is somehow known and we want to subsequently measure the positions of the particles, this measurement process itself will unavoidably modify the initial wave function.

The second difference between classical and quantum theories has to do with the typical initial conditions that can be assumed in scenarios where their exact values cannot be known. Most quantum experiments are analyzed by assuming  the initial wave function of the system to be given (say, by means of a suitable procedure of preparation of the system).  Thus, in order to analyze quantum experiments, only a statistical assumption  regarding the distribution of the initial positions of the particles  is needed. The so-called quantum equilibrium hypothesis states  that the distribution of the positions of particles is given by the modulus squared of their wave function \cite{Durr1992}. This hypothesis is of paramount importance, since it allows us to quantify precisely the the degree of quantum uncertainty in an experiment and it guarantees the empirical equivalence of Bohmian mechanics and the standard quantum-mechanical approach. 

Now that we have given this brief introduction to the topic, we can advance that the structure of the rest of the paper is as follows. In section \ref{sec2}, we discuss why, according to Bohmian mechanics, the initial conditions cannot be properly measured. First, in subsection \ref{sec21}, we briefly introduce the dynamical laws of the theory and we discuss the distinction between the universal wave function, the conditional wave function and the effective wave function of a subsystem; since these distinctions are important for understanding the origin of uncertainties in Bohmian mechanics. Then in subsection \ref{sec22}, after explaining why in classical mechanics the initial conditions can in principle be measured, we show why the wave function cannot be measured; and finally, we show that a measurement of position typically disturbs the wave function. In section \ref{sec3}, we introduce the quantum equilibrium hypothesis, explaining its meaning in subsection \ref{sec31}, its justification in subsection \ref{sec32} and in subsection \ref{sec33}, the absolute unpredictability of quantum mechanics that emerges as a direct consequence of quantum equilibrium hypothesis. Next, in section \ref{sec4}, we add some further considerations concerning the relation between randomness, measurement and equilibrium. Then finally, in section \ref{sec5}, we summarize our conclusions.

\section{A dynamical discussion of randomness}
\label{sec2}

In this section, we want to highlight the differences between classical measurements and Bohmian measurements. As we have just mentioned, we will show that, according to Bohmian mechanics, there are fundamental limits to the possibility of determining the initial conditions of a given system through measurement---a sort of limit that we do not find in classical measurements. Due to this inherent limitation on the determination of the initial conditions, the subsequent evolution of the system cannot be predicted. 

\subsection{A brief introduction to the dynamics of Bohmian mechanics}
\label{sec21}

Throughout this section we will be mainly concerned with the dynamics of Bohmian Mechanics. So, in what follows we present the basic dynamical laws of the theory\cite{durr13,Oriols12}. We take advantage of this presentation to introduce three different types of (Bohmian) wave functions that are relevant for our later discussion. For ease of exposition and understanding, we will focus on a non-relativistic Bohmian world with spinless particles \lq\lq{}living\rq\rq{} in a one-dimensional physical space. The generalization to a 3D physical space inhabited by particles with spin does not change any of our conclusions. 

\subsubsection{The universal wave function}
\label{sec211}

Bohmian Mechanics is a quantum theory in which the complete state of the whole system of particles in the universe, say $N$, is given by the \emph{universal} wave function $\Psi(\mathbf{x},t)$ with $\mathbf{ x}=\{x_1,x_2,...,x_N\}$, and by the actual positions of the particles $\mathbf{X}(t) = \{ X_1(t),X_2(t),...,X_N(t)\}$. The universal wave function evolves according to the many-particle Schr\"odinger equation:
\begin{eqnarray}
i \hbar \frac{\partial \Psi(\mathbf{x},t)}{\partial t} =\hat H  \Psi(\mathbf{x},t) = \left(  \sum_{j = 1}^N  -\frac{\hbar^2}{2m_i} \frac{\partial^2}{\partial x_j^2}+ U(\mathbf{x},t) \right) \Psi(\mathbf{x},t) ,
\label{scho}
\end{eqnarray}
where $m_i$ is the mass of the $i$-th particle  and the Hamiltonian operator $\hat H$ contains, both, the kinetic energy term and the potential energy term, $U(\mathbf{x},t)$, of all the particles. In addition, the trajectory of each particle is given by the integral of the particle\rq{}s velocity $v_i(\mathbf{x},t)$ defined through the so-called \emph{guidance equation}:
\begin{equation}
v_i(\mathbf{x},t)\Big|_{\mathbf{x} = \mathbf{X}(t)}=\frac{d X_i(t)}{dt} = \frac{\hbar}{m_i} \text{Im} \left(\frac {1}{\Psi(\mathbf{x},t)}  \frac{\partial \Psi(\mathbf{x},t)} {\partial x_i}  \right) \Big|_{\mathbf{x} = \mathbf{X}(t)}.
\label{velo}
\end{equation}
In order to derive the Bohmian trajectories from Eq. (\ref{velo}), we need to specify the initial position of all the particles, $\mathbf{X}(0) = \{ X_1(0),X_2(0),...,X_N(0)\}$. In turn, in order to solve the Schr\"odinger equation given in Eq. (\ref{scho}), the universal wave function at a given initial time, $\Psi(\mathbf{x},0)$, must be specified. As we say above in the Introduction, Bohmian mechanics is a completely deterministic theory in the sense that specification of the complete state of the universe at one initial time, $\big(\Psi(\mathbf{x},0),\mathbf{X}(0)\big)$, together with the laws given in Eqs. (\ref{scho}) and (\ref{velo}), is compatible with just one state of the universe for any other time $t$, $\big(\Psi(\mathbf{x},t),\mathbf{X}(t)\big)$. In this regard, the dynamical laws of Bohmian mechanics are as deterministic as the dynamical (Newtonian or Hamiltonian) laws of classical mechanics\footnote{We should mention that there is no measurement postulate in the Bohmian theory because, in this theory, the system-apparatus interaction is regarded in the same way as any other type of (deterministic) interaction.}. 

The following continuity equation for the modulus squared of the wave function can be straightforwardly obtained from Eq. (\ref{scho}):
\begin{eqnarray}
\label{continuity}
\frac{\partial |\Psi(\mathbf{x},t)|^2}{\partial t}+\sum_{i=1}^{N}  \frac{\partial } {\partial x_i} \big( |\Psi(\mathbf{x},t)|^2 v_i(\mathbf{x},t) \big)=0.
\end{eqnarray}
Now, suppose that we do not know the initial positions of the particles, but that we assume that they are in accordance with a given statistical distribution $\rho(\mathbf{x},0)$. It follows from Eq. (\ref{continuity}) that if this initial distribution is given by $\rho(\mathbf{x},0)=|\Psi(\mathbf{x},0)|^2$, then the dynamics preserves the form of this distribution; that is, for any other future time $t$, the distribution is $\rho(\mathbf{x},t)=|\Psi(\mathbf{x},t)|^2$. When the distribution is preserved by the dynamics, as happens for $\rho(\mathbf{x},0)=|\Psi(\mathbf{x},0)|^2$, we say that this distribution is equivariant.

\subsubsection{The conditional wave function}
\label{sec212}

In situations of practical interest, we are not concerned with the whole universe but with smaller subsystems, i.e., a particle or a collection of particles in our laboratory. However, the wave function that appears in the postulates of Bohmian mechanics, Eqs. (\ref{scho}) and (\ref{velo}), is the universal wave function, $\Psi(\mathbf{x},t)$. If we want the theory to make contact with real applications, we need some tools that enable us to refer to and to characterize arbitrary subsystems of the universe. In Bohmian mechanics, those tools are the so-called \emph{conditional} and \emph{effective} wave function of a given system. 

Let us first define the \emph{conditional} wave function of a system. For simplicity, we will focus on a specific particle, labeled $a$, with the configuration variable $x_a$\footnote{The generalization to a conditional wave function with many degrees of freedom can be derived straightforwardly.}. By definition, the conditional wave function $\psi_{a}(x_a,t)$ is: 
\begin{equation}
\psi_{a}(x_a,t) = \Psi(x_a,\mathbf{Y}(t),t),
\label{eq_conditional}
\end{equation}
where $\mathbf{Y}(t) =\{X_1(t), ..., X_{a-1}(t),X_{a+1}(t),...,X_{N}(t)\}$ are the positions at $t$ of all the particles except $a$. It is usual to refer to the set of all particles except $a$ as the environment of $a$. Note that whereas the universal wave function is a function defined in the $N$-dimensional configuration space of the universe, the conditional wave function is a function only of $x_a$ (and time). 
    It follows from Eq. (\ref{velo}), that the velocity of $a$ is given by the function $\psi_a(x_a,t)$ through the corresponding guidance equation:
\begin{eqnarray}
v_a(\mathbf{x},t)\Big|_{\mathbf{x}=\mathbf{X}(t)} &=& \frac{\hbar}{m_a} \text{Im} \left(\frac {1}{\Psi(\mathbf{x},t)}  \frac{\partial \Psi(\mathbf{x},t)} {\partial x_a}  \right)\Big|_{\mathbf{x} = \mathbf{X}(t)}\nonumber\\ &=&  \frac{\hbar}{m_a} \text{Im} \left(\frac {1}{\psi_a(x_a,t)}  \frac{\partial \psi_a(x_a,t)} {\partial x_a}  \right)\Big|_{x_a = X_a(t)}.
\label{eq-guid-cwf}
\end{eqnarray}
Therefore, all the dynamics of a quantum (sub)system can be inferred  from its conditional wave function. However, in general, the conditional wave function does \emph{not} evolve according to the Schr\"odinger equation, but it obeys its own (more complicated) equation \cite{Oriols2007}:
\begin{eqnarray}
\label{eq_conditional}
i\hbar\frac{\partial \psi_a(x_a,t)}{\partial t}&=&\Big(-\frac{\hbar^2}{2m}\frac{ \partial}{\partial x_a^2}+U(x_a,\mathbf{Y}(t),t)\nonumber\\ &+&G(x_a,\mathbf{Y}(t),t)+ i J(x_a,\mathbf{Y}(t),t) \Big) \psi_{a}(x_a,t),
\end{eqnarray}
where $U(x_a,\mathbf{Y}(t),t)$ is the potential that appears in Eq. (\ref{scho}) evaluated at $\mathbf{Y}(t)$. Eq. (\ref{eq_conditional}) demonstrates that such a single-particle wave equation exists; however, we do not know the exact form of the terms $G(x_a,\mathbf{Y}(t),t)$ and $J(x_a,\mathbf{Y}(t),t)$ because to know these terms we would need to have complete knowledge of the (universal) wave function \cite{Oriols2007}. Our ignorance of these terms is certainly a source of uncertainty in our predictions. This source of uncertainty is of the sort that appears in (classical or quantum) open systems due to the interchange of particles and energy with the (unknown) environment. 

\subsubsection{The effective wave function}
\label{sec213}

While the evolution of the conditional wave function for a quantum subsystem is given by Eq. (\ref{eq_conditional}), which is not exactly the Schr\"odinger equation, Bohmian theory provides the concept of the \emph{effective} wave function of a quantum subsystem, whose equation of motion is exactly the Schr\"odinger equation. Once again, let us focus on particle $a$ and let $\mathbf{y} \equiv (x_1,....,x_{a-1},x_{a+1},...,x_N)$ be the position variables of $a$'s environment\footnote{The generalization to an effective wave function with many degrees of freedom can be derived straightforwardly.}. Now suppose that the universal wave function, over some time interval (for example, for $t \in [t_0,t_1]$), can be decomposed in the form:
\begin{equation}
\Psi(x_a,\mathbf{y},t) = \phi_a(x_a,t) \theta(\mathbf{y},t) +\Psi^{\dagger}(x_a,\mathbf{y},t)
\label{effective}
\end{equation}	 
where $\theta(\mathbf{y},t)$  and  $\Psi^{\dagger}(x_a,\mathbf{y},t)$ are functions with macroscopically disjoint supports. In addition, we assume that $\mathbf{Y}(t)$ lies within the support of $\theta(\mathbf{y},t)$. If these conditions are met, then we refer to $\phi_a(x_a,t)$  as the effective wave function of $a$. It is straightforward to show that the equation of motion for the effective wave function $\phi_a(x_a,t)$ during $t \in [t_0,t_1]$ is the usual Schr\"odinger equation:
\begin{eqnarray}
\label{eq_effective}
i\hbar\frac{\partial \phi_a(x_a,t)}{\partial t}=\Big(-\frac{\hbar^2}{2m}\frac{ \partial}{\partial x_a^2}+U_a(x_a,t)\Big) \phi_{a}(x_a,t),
\end{eqnarray}
and the Bohmian velocity is just:
\begin{eqnarray}
v_a(x_a,t)\Big|_{x_a = X_a(t)}=  \frac{\hbar}{m_a} \text{Im} \left(\frac {1}{\phi_a(x_a,t)}  \frac{\partial \phi_a(x_a,t)} {\partial x_a}  \right)\Big|_{x_a = X_a(t)}.
\label{velo_eff}
\end{eqnarray}
We should note that the form of the universal wave function $\Psi(x_a,\mathbf{y},t)$ in Eq. (\ref{effective}) implies a separable potential $U(\mathbf{x},t)=U_a(x_a,t)+U_y(\mathbf{y},t)$. We emphasize that the effective wave function of a quantum (sub)system does not always exist; however, when the effective wave function exists, it is equal to the conditional wave function\footnote{This can be seen if one makes the substitution $\mathbf{y}=\mathbf{Y}(t)$ in Eq. (\ref{effective}) and recalls that, by assumption, $\mathbf{Y}(t)$ lies within the support of $\theta(\mathbf{y},t)$. }. 

In exactly the same way as we developed Eq. (\ref{continuity}), a new continuity equation can be deduced from Eq. (\ref{eq_effective}):
\begin{eqnarray}
\label{continuity2}
\frac{\partial |\phi_a(x_a,t)|^2}{\partial t}+ \frac{\partial } {\partial x_a} \big( |\phi_a(x_a,t)|^2 v_a(x_a,t) \big)=0.
\end{eqnarray}
Whenever in orthodox quantum mechanics a definite wave function is attributed to a given subsystem, that wave function corresponds with the Bohmian effective wave function. We will see in subsection \ref{sec31} that Eq. (\ref{continuity2}) guarantees equivariance, which will be relevant to ensure the empirical equivalence between Bohmian and orthodox quantum mechanics.

Bohmian mechanics is a holistic theory. Strictly speaking, its postulates Eqs. (\ref{scho}) and (\ref{velo}) only apply to the universe as a whole. Therefore, when it comes to assessing the ontology of the theory, the fundamental objects are the particles with positions $\mathbf{X}(t) = \{ X_1(t),X_2(t),...,X_N(t)\}$ and the universal wave function $\Psi(\mathbf{x},t)$.  As we have seen, in order to deal with subsystems of the universe, the conditional wave function is introduced. The conditional wave function of a given subsystem is not fundamental but it supervenes on the fundamental $\mathbf{X}(t)$ and $\Psi$.  From Eq. (\ref{eq-guid-cwf}), it can be seen that the complete dynamics of a subsystem is determined once both its conditional wave function and the initial positions of its constituent particles are specified. Thus, we can say that the complete state of a subsystem is given by its conditional wave function and the position of its particles---even though its conditional wave function is not a primitive object in Bohmian mechanics. 

The conditional wave function for a subsystem does not evolve according to the Schr\"odinger equation, but is a solution of the more complicated Eq. (\ref{eq_conditional}). Yet we have seen that, when a subsystem is sufficiently decoupled from its environment so that the conditions for it to have a well-defined effective wave function are satisfied, then its effective wave function actually does obey the Schr\"odinger equation, Eq. (\ref{eq_effective}). In this case, once again the evolution of the system is determined when both the initial positions of its constituent particles and its effective wave function are specified. In the rest of the paper, when we refer to the wave function of a given subsystem of the universe, we are of course referring to its conditional wave function (or to its effective wave function, if the system has one) even if we do not make it explicit.

\subsection{Explanation of unpredictability based on the dynamics}
\label{sec22}

We next want to highlight the contrast between classical and Bohmian mechanics when it comes to experimentally determining the initial conditions of a system. In order to do so, in subsection \ref{sec221} we discuss a very simple model of a classical measurement interaction represented by an impulsive Hamiltonian. It will follow from this discussion that, in principle, both the position and the velocity of a particle can be measured without an appreciable perturbation. Then, in subsection \ref{sec222}, we show why the wave function can\textit{not} be measured; and in subsection \ref{sec223}, we show that a Bohmian measurement of the position typically disturbs the wave function. 

\subsubsection{Measurement of the initial conditions in classical mechanics}
\label{sec221}

We consider a one-dimensional classical system (object) of mass $m$ and the measurement of the property $A (x, p_x)$ of the system, where $x$ is the position coordinate of the object and $p_x$ its momentum. To perform the measurement, we consider a pointer, whose position and momentum are denoted by $y$ and $p_y$ respectively. We assume that the Hamiltonian governing the interaction between the object and the apparatus has the following form:
\begin{eqnarray}
\label{al1}
H_{int}=g A(x,p_x) p_y,
\end{eqnarray}
where $g$ is a coupling constant. The total Hamiltonian of the system-plus-apparatus is composed of the interaction term $H_{int}$ in Eq. (\ref{al1}) plus the kinetic energy of the object and of the apparatus. We consider that during the course of the interaction $H_{int}$ is the only relevant term in the Hamiltonian. Then, the classical equations of motion can be obtained by substituting Eq. (\ref{al1}) into Hamilton\rq{}s equations:
\begin{eqnarray}
\label{al2}
\frac{dx}{dt}&=&\frac {\partial H_{int}}{\partial p_x}=g \frac{\partial A(x,p_x)}{\partial p_x} p_y,\\
\label{al3}
\frac{dy}{dt}&=&\frac {\partial H_{int}}{\partial p_y}=g A(x,p_x)\\
\label{al4}
\frac{d p_x}{dt}&=&-\frac {\partial H_{int}}{\partial x}=-g \frac{\partial A(x,p_x)}{\partial x} p_y,\\
\label{al5}
\frac{d p_y}{dt}&=&-\frac {\partial H_{int}}{\partial y}=0
\end{eqnarray}
It can be seen from Eq. (\ref{al3}) that the value of the time derivative of the  position of the apparatus is correlated with the value of $A(x, p_x)$. From Eqs. (\ref{al2}) and  (\ref{al4}), we see, however, that the variables $x$ and $p_x$ are disturbed during the measurement process.  However, such disturbance can be made arbitrarily small by assuming $p_y \approx 0$. From  Eq. (\ref{al5}) it follows that the  momentum of the pointer remains constant during the whole interaction (its temporal derivative is zero). Therefore, if we consider $P_y(0) \approx 0$ at the initial time, the expressions  Eqs. (\ref{al2}) and (\ref{al4}) are greatly simplified. In this case, the above equations can be easily integrated, yielding:
\begin{eqnarray}
\label{al6}
\label{al9b}
X(t) &=& X(0) \;\; ; \;\;\;\;\; Y (t) = Y(0) + gA (X(0), P_x(0)) t\\
P_x (t) &=& P_x(0)\;\; ;\; \;\;  P_y (t) = P_y(0)
\end{eqnarray}
Now it can be seen that neither the position of the object nor its momentum are altered during the measurement process. We would like to note that from the second expression in Eq. (\ref{al9b}), the value of $A (X(0), P_x(0))$ can be determined if the initial and final positions of the pointer, $Y (0)$ and $Y (t)$, are known. 

Given these results, it is easy to show how we can measure the position and momentum of the object at the same time without causing a significant disturbance their values. To do this, we just assume an interaction with two pieces of apparatus (whose pointers are represented on this occasion by the variables $y$, $p_y$ and $z$, $p_z$, respectively) so that the  interaction Hamiltonian is now:
\begin{eqnarray}
\label{al8}
H_{int}=g x p_y + h p_x p_z,
\end{eqnarray}
where $g$ and $h$ are the coupling constants between the object and the first and second pieces of apparatus, respectively. Applying simplifications analogous to those in the previous case, the following equations of motion are obtained:
\begin{eqnarray}
\label{al9}
X (t) &=& X(0) \;\; ; \;\;\;\;\; Y (t) = Y(0) + X(0)  t \;\; ; \;\;  Z (t) = Y(0) + P_x(0)  t\\
P_x (t) &=& P_x(0)\;\; ;\; \;\;  P_y (t) = P_y(0)\;\; ; \;\;\;\;\;\;\;\;\;\;  P_z (t) =P_z(0)
\end{eqnarray}
If the position of the respective pointers at the end of the interaction is known, then both the position and the initial velocity of the object can be directly inferred from Eq. (\ref{al9}). Having this knowledge is a necessary (but not sufficient) condition for precisely predicting the posterior behavior of the particle. As we will see next, in Bohmian mechanics the detailed initial conditions of a system cannot be properly measured, so in the context of this theory not even this necessary condition can be met.

\subsubsection{Measurement of the wave function in quantum mechanics}
\label{sec222}

As we have repeatedly stressed, the initial conditions of a quantum system are its wave function $\psi$ and the initial positions of its constituent particles. As opposed to what happens in the classical case, in Bohmian mechanics one of the initial conditions---the wave function---cannot be properly measured \cite{clone}. The argument leading to the non-measurability of the wave function is rather simple and only the linearity of the quantum dynamical laws needs to be assumed.

We are interested in developing an apparatus that is capable of measuring the wave function $\psi(x,t)$ of an object with position variable $x$. We will assume that such measuring apparatus is an additional quantum system whose final pointer position $Y(t_f)$ allows us to infer the wave function that is subjected to the measurement. Let $\phi_R(y,0)$ be the (effective) wave function of the apparatus when it is in the initial state---ready for the measurement---where $y$ is the position variable of the apparatus. When needed, we will refer to the (effective) wave function of the composite system as $\Phi(x,y,t)$. We consider, for the sake of our argument, two possible initial wave functions of the object, $\psi_1(x,0)$ and $\psi_2(x,0)$, which allows us to define two different initial wave functions of the composite system, $\Phi_1(x,y,0)=\psi_1(x,0)\phi_R(y,0)$ and $\Phi_2(x,y,0)=\psi_2(x,0) \phi_R(y,0)$.  A proper measurement interaction must provoke a correlation between the wave function of the system and the pointer of the apparatus, so that the two initial wave functions, $\Phi_1(x,y,0)$ and $\Phi_2(x,y,0)$, are associated with two different values of the pointer's position at the end of the measurement. The temporal evolution of the joint wave function $\Phi(x,y,t)$ leading to the desired correlation between the object and the apparatus is given by the linear Schr\"odinger equation: 
\begin{eqnarray}
i \hbar \frac{\partial \Phi(x,y,t)}{\partial t} = \left(  -\frac{\hbar^2}{2m_x }\frac{\partial^2}{\partial x^2}-\frac{\hbar^2}{2m_y} \frac{\partial^2}{\partial y^2}+  B(x,y,t) \right) \Phi(x,y,t).
\label{schobis2}
\end{eqnarray}
where the term $B(x,y,t)$ is responsible for the proper correlation between any $\psi(x,t)$ and $Y(t)$. Note that the evolution of $\Phi(x,y,t)$ during the measurement is not separable. 

In order for the interaction to constitute an (ideal) measurement of the wave function of the object, the evolution of the global wave function from $t=0$ till $t=t_f$, through the corresponding Schr\"odinger equation in (\ref{schobis2}), must be:
\begin{eqnarray}
\Phi_1(x,y,0)=\psi_1(x,0)\phi_R(y,0) \rightarrow \Phi_1(x,y,t_f)=\psi_1(x,t_f)\phi_1(y,t_f) \label{1} 
\end{eqnarray}
where  $\phi_1(y,t_f)$ is the final wave function of the apparatus indicating that the measured wave function is $\psi_1(x,0)$. In the same way: 
\begin{eqnarray}
\Phi_2(x,y,0)=\psi_2(x,0)\phi_R(y,0) \rightarrow \Phi_2(x,y,t_f)=\psi_2(x,t_f)\phi_2(y,t_f) \label{2}
\end{eqnarray}
where  $\phi_2(y,t_f)$ is the final wave function of the apparatus indicating that the measured wave function is $\psi_2(x,0)$. The wave functions $\phi_1(y,t_f)$  and  $\phi_2(y,t_f)$ need to have macroscopically disjoint supports\footnote{We define the support of $\psi$,  $supp\; \psi$,  as the set of all points of its domain such that the value of the wave function $\psi$ is significantly different from zero.} so that when we observe $Y(t_f) \in supp\; \phi_1(y,t_f)$, we take this as an unambiguous indication that the result of the measurement is that the wave function of the object is $\psi_1(x,0)$. Similarly, when we observe $Y(t_f) \in supp\; \phi_2(y,t_f)$, we take this as an unambiguous indication that the result of the measurement is that the wave function of the object is $\psi_2(x,0)$.

   Now, let the initial wave function to be measured be $\psi_3(x,0)=(\psi_2(x,0)+\psi_1(x,0))/\sqrt{2}$. Note that such a state always exists in the Hilbert space of the object system. If we use the same measuring device as before, it follows from (\ref{1}) and (\ref{2}), and the linearity of the Schr\"odinger equation that the evolution of the joint state will be:
\begin{eqnarray}
\Phi_3(x,y,0)=\frac{ \psi_2(x,0)+\psi_1(x,0) } {\sqrt{2}} \phi_R(y,0) \rightarrow \frac{\psi_1(x,t_f)\phi_1(y,t_f)}{\sqrt{2}}+\frac{\psi_2(x,t_f)\phi_2(y,t_f)}{\sqrt{2}}\nonumber\\ \label{3} 
\end{eqnarray}
Given the final state $\psi_1(x,t_f)\phi_1(y,t_f)/\sqrt{2}+\psi_2(x,t_f)\phi_2(y,t_f)/\sqrt{2}$, the laws of Bohmian mechanics entail that, at the end of the measurement, the pointer's position will lie either within the support of $\phi_1(y,t)$ or within the support of $\phi_2(y,t)$. However, if $Y(t_f) \in supp\; \phi_1(y,t_f)$, we have assumed that the result of the measurement is that the wave function is $\psi_1(x,t_f)$ (and not $\psi_3(x,t_f)$); alternatively, if $Y(t_f) \in supp\; \phi_2(y,t_f)$, the pointer's position is taken as an indication that the result of the measurement is that the wave function is $\psi_2(x,t_f)$ (and not $\psi_3(x,t_f)$).
   In conclusion, it can be seen that it is not possible to have an apparatus whose operation is based on the linearity of the Schr\"odinger equation that measures all wave functions\footnote{Any additional information regarding a particular $\psi$ could imply the inclusion of a $\psi$-dependent term $B(x,y,t,\psi)$ which breaks the linearity of Eq. (\ref{schobis2}).}. If our measuring apparatus is designed to measure the wave functions $\psi_1(x,t_f)$ and $\psi_2(x,t_f)$, it will be unable to measure correctly a new wave function constructed as a linear combination of the previous two, such as  $\psi_3(x,0)=(\psi_2(x,0)+\psi_1(x,0))/\sqrt{2}$.

\subsubsection{Measurement of position in quantum mechanics}
\label{sec223}

In the previous subsection, we show that the wave function cannot be measured. Since the motion of a quantum system is governed by the wave function, this has important consequences for the unpredictability of the future behavior of quantum systems. At this point, the reader may object that, in many situations, physicists assume that they know what the wave function of a given system is. For instance, by measuring the energy of a system, we can assume that the effective wave function of the system (after the measurement) is an eigenstate corresponding to the measured energy eigenvalue. However, strictly speaking, this procedure for obtaining information on the wave function cannot be considered a measurement of the wave function, since the wave function has not been measured; only the energy has\footnote{The wave function is a hidden variable in the sense that it cannot be measured directly and we can only obtain information about it from the measurement of other variables. It is surprising that, nevertheless, the expression ``hidden variable'' has typically been reserved for the positions (and not the wave function) when in fact the positions can be routinely measured just by using a suitably placed detecting screen. }. In the literature, this procedure is defined as ``preparation'' of the wave function. 

In this subsection, we bolster our results from the previous subsection by showing that, if we assume that we know the initial wave function thanks to a specific preparation technique, and we want to measure the Bohmian position of the particle in the system in order to determine both initial conditions, the measurement of the position perturbs the initial wave function so we cannot have knowledge of both. 

\begin{figure}
\includegraphics[width=0.824\columnwidth]{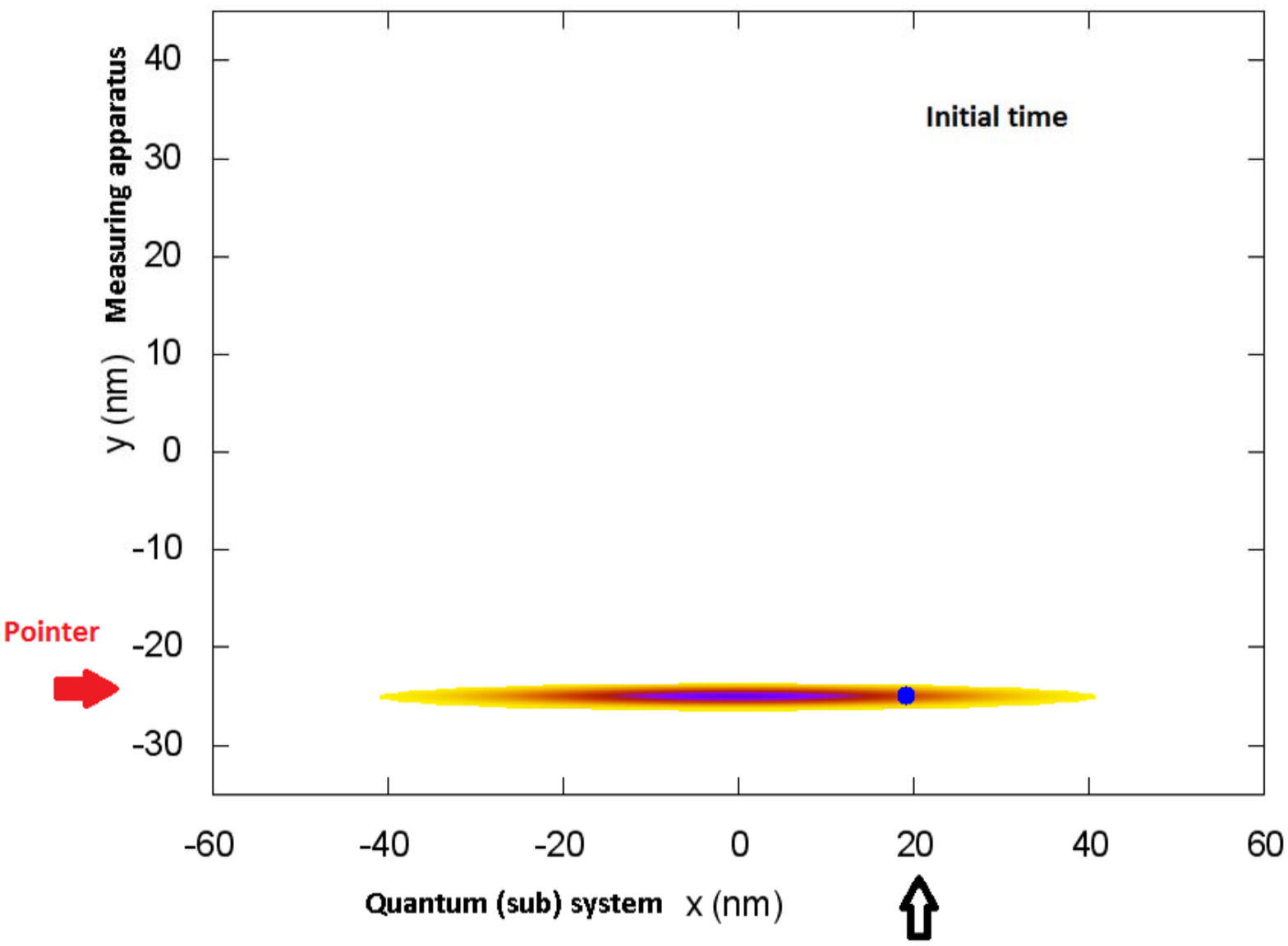}
\caption{Probability density of the initial (effective) wave function  $\Phi(x,y,0)$ and (filled-in (blue) circle) the corresponding (Bohmian) position $\{X(0),Y(0)) \}$ in the two-dimensional system-plus-apparatus  configuration space. (Color figure online).}
\label{fig1}
\end{figure}

Let us consider, again, that we have two systems: the object with position variable $x$ and the apparatus with position variable $y$. We can assume that this latter position represents the center of mass of the pointer. We have prepared the object system in such a way that we know its wave function and we subsequently want to measure its position. The (effective) two-dimensional wave function of system plus apparatus is $\Phi(x,y,t)$ and its initial value is, for example, the product of two Gaussian wave functions at rest with initial dispersions of $\sigma_x=15\:nm$ and $\sigma_y=0.5\:nm$ respectively:
\begin{equation}
\Phi(x,y,0) = \frac{1}{\sqrt{\sigma_x\sqrt{\pi}}} e^{ -\frac{(x-x_0)^2}{2 \sigma_x^2} } \frac{1}{\sqrt{\sigma_y\sqrt{\pi}}} e^{ -\frac{(y-y_0)^2}{2 \sigma_y^2} }.
\label{eq:gauss}
\end{equation} 
The modulus at the initial time, $|\Phi(x,y,0)|^2$, together with the initial positions of the system and pointer, $\{X(0),Y(0) \}$, is plotted in Fig. \ref{fig1}. 

\begin{figure}
\includegraphics[width=0.824\columnwidth]{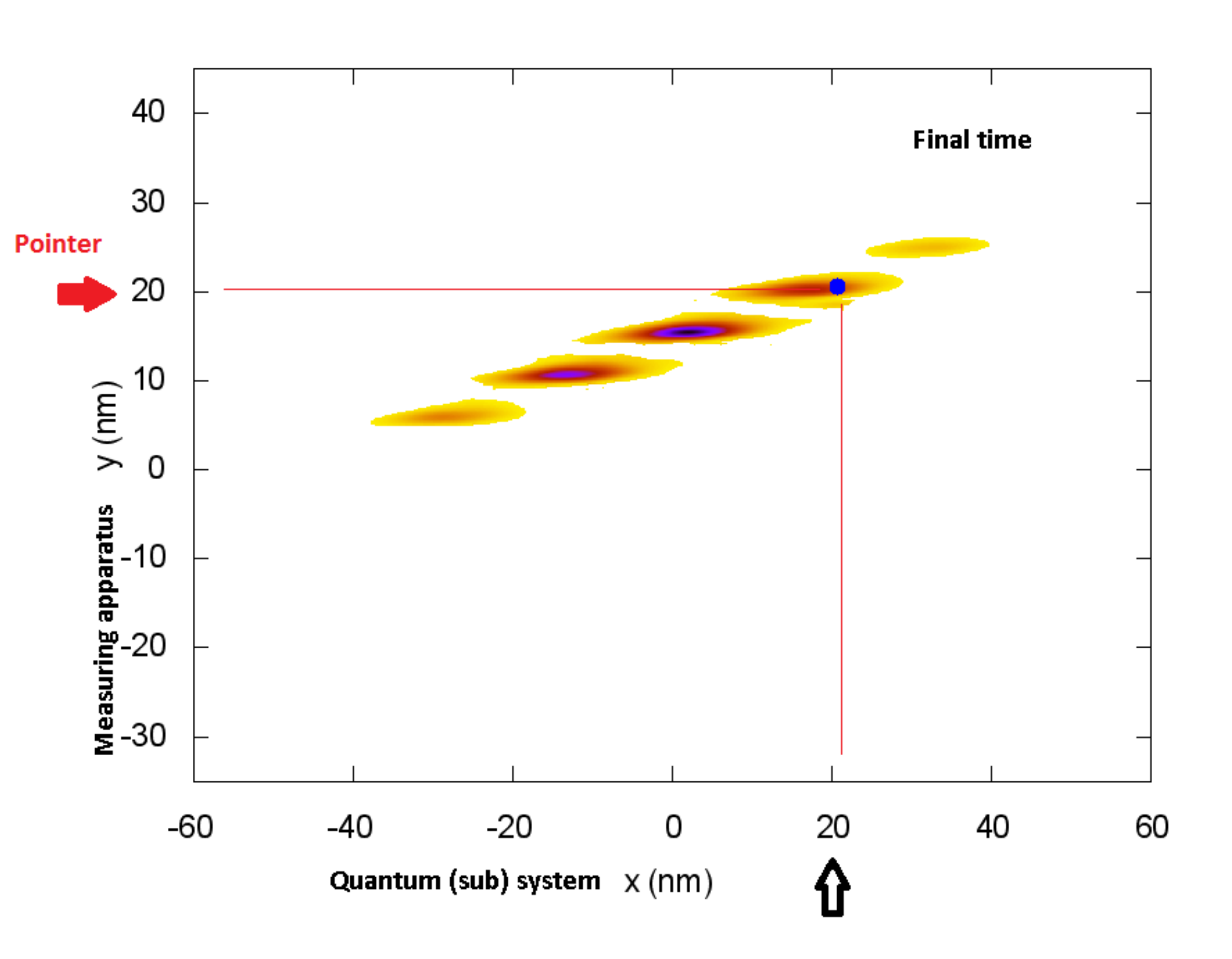}
\caption{Probability density of the (effective) wave function  $\Phi(x,y,t_f)$ at the final time $t_f$ and (filled-in (blue) circle) the corresponding (Bohmian) position $\{X(t_f),Y(t_f)) \}$ in the two-dimensional system-plus-apparatus  configuration space. The (effective) unitary Schr\"odinger equation of the system plus apparatus does not depend on the Bohmian position, but its evolution allows a correlation between different $X(t)$ and different $Y(t)$. Such correlation is a  measurement of the Bohmian position of the particle, with the corresponding perturbation of the wave function (Color figure online).}
\label{fig2}
\end{figure}

We will assume that the interaction is governed by the following interaction Hamiltonian $H_{int}=g A(x) p_y$, similar to the Hamiltonian we use in Eq. (\ref{al1}) for modelling a classical measurement. Here, $g=10^{5}\;m/s$, $p_y=i \hbar \partial /\partial y$  and the function $A(x)=2n_p(x)+1$ where $n_p(x)$ is equal to the integer number $p$ when $p \Delta \leq x < (p+1)\Delta$ with $\Delta=15\:nm$.  In the limit $\Delta \rightarrow 0$ we have $A(x) \propto x$ as in  Eq. (\ref{al8}). Therefore, in our simulation, we use the following equation:
\begin{eqnarray}
i \hbar \frac{\partial \Phi(x,y,t)}{\partial t} = \left(  -\frac{\hbar^2}{2m_x }\frac{\partial^2}{\partial x^2}-\frac{\hbar^2}{2m_y} \frac{\partial^2}{\partial y^2}+  g  A(x) i \hbar \frac{\partial}{\partial y} \right) \Phi(x,y,t).
\label{schobis}
\end{eqnarray}
The interacting term $H_{int}$ generates a correlation between the position of the system $X(t)$ and the position of the pointer $Y(t)$. Such a correlation allows us to \rq\rq{}look\rq\rq{} at the position of the pointer and infer from it the position of the system (with some technical uncertainty related to the finite value of $\Delta$, which would disappear in the limit $\Delta \rightarrow 0$). In Fig. \ref{fig2}, the pointer of the apparatus indicates a final value $Y(t_f)=20\; nm$ at the final time $t_f$ after the measurement, which is perfectly correlated with the position of the Bohmian particle $X(t_f)=20\; nm$. At this final time, the one-dimensional (effective or conditional) wave function of the quantum system alone can be defined as $\phi(x,t_f)=\psi(x,t_f)\equiv \Phi(x,Y(t_f),t_f)$. The relevant point is that the one-dimensional (conditional) wave function $\Phi(x,Y(t_f),t_f)$ strongly depends on the pointer position and, even worse, it does not resemble the initial conditional wave function $\Phi(x,Y(0),0)$ at all. The moral of this analysis should now be clear: measuring the position of the system involves a great perturbation of its wave function.  

It is important to note that if we use a highly localized wave function in the x-direction, instead of the Gaussian wave packet in Eq. (\ref{eq:gauss}), then we will obtain information on the position without disturbing this localized wave packet. Here, we are referring to a wave packet whose support is localized inside a step interval $\Delta$ (i.e., eigenfunctions of the position operator when $\Delta \rightarrow 0$). Then, for this localized wave packet, the term $A(x)$ has no spatial dependence and $H_{int}$ only affects the $y$ wave function. In other words, Eq. (\ref{schobis}) is separable and the measurement of the position does not perturb the (localized) wave function. Unfortunately, most of the wave functions of practical interest are not  eigenfunctions of the position operator and measuring the position implies perturbation of the wave function. It should be noted, in addition, that the time evolution of a spatially extremely narrow wave function (close to a delta function), due to the guidance equation (\ref{velo}), implies such a large dispersion in the velocity of the particle that an infinitesimal variation of the initial position inside the initial wave packet provokes a large variation in its final position (as in a chaotic system \cite{chaos}). In short, if the wave function is close to a position eigenstate, we are in the situation of approximately knowing the wave function and the position at the initial time through a measurement; however, this situation also leads to an effective unpredictability of the future behavior of the system.  

So far we have shown that, assuming a particular Hamiltonian interaction, $H_{int}=gA(x)p_y$, a measurement of the position disturbs the initial conditional wave function (unless the latter is a position eigenstate). Obviously, this is not a general result and the reader may wonder whether, by means of another interaction between particles $x$ and $y$, the position of $x$ can be measured without disturbing the wave function. We want to argue, next, that this cannot be the case.

In order to do so, we will assume an interaction between $x$ and $y$ such that it does not provoke a modification of the conditional wave function of $x$\footnote{In order to avoid any perturbation in the x-wave function (as discussed for the position eigenstate), we consider $\Delta \rightarrow \infty$ which means that we have eliminated the x-dependence of the function $A$ in Eq. (\ref{schobis}). Therefore, from Eq. (\ref{eq:gauss}), the x-wave packet evolves independently of the y-wave packet at all times. One can clearly see that, as desired, the velocity in the x-direction is zero because the phase of the Gaussian in the x-direction remains x-independent all the times: $v_x(x,y,0)=v_x(x,0)=  \frac{\hbar}{m} \text{Im} \left(\frac {1}{\Phi(x,y,0)}  \frac{\partial \Phi(x,y,0)} {\partial x}  \right)=0$}. If this interaction is to be a measurement of the position, then obviously it should connect different initial positions of $x$ with different final positions of the pointer, so that by looking at the final position of the pointer $y$, the position of $x$ can be inferred. It is clear then that a necessary condition for the interaction to constitute a measurement is that it needs to induce a “channelization” of $\Phi(x,y,t_f)$ in the $y$-space so that at the final time $t_f$, several positions of the pointer can be discriminated. Now, if this latter requirement is combined with the former (namely, that the conditional wave function of $x$ is not disturbed as a result of the interaction), we obtain a situation similar to that depicted in Fig. \ref{fig3}. Here, we see that the different branches of $\Phi(x,y,t_f)$ associated with the different possible final positions of the pointer all have the same shape in the $x$-space. Note, that this is in striking contrast with the situation depicted in Fig. \ref{fig2} where each branch of  $\Phi(x,y,t_f)$ associated with one possible final position of the pointer clearly has a different projection on the $x$-axis. But the former result is imposed by the requirement that, for all possible final positions of the pointer $Y_i(t_f)$, the conditional/effective wave function $\Phi(x,Y_i(t_f),t_f)$ must be equal to the initial effective wave function of $x$: $\Phi(x,Y(0),0)$.

The evolution represented in Fig. \ref{fig3} has a dramatic consequence. It does not produce any dynamical correlation between $Y(t)$ and $X(t)$. It can be accepted that, by chance, the pointer may indicate $Y (t_f ) = 20$ nm when the position of the system is $X(t_f ) = 20$ nm. Yet there will be no such coincidence for other possible initial positions of the pointer. In fact, according to the quantum equilibrium hypothesis, which we discuss in more detail in the next section, there are infinitely many Bohmian trajectories such that, while the pointer ends in position $Y(t_f ) = 20$ nm, the position of $x$ is not $X(t_f ) = 20$ nm but any other value. So it is clear that the interaction represented in Fig. \ref{fig3} cannot be considered a measurement of $x$ position.

A final remark is required here. We can conceive a rather special universe in which, for all systems initially characterized by the wave function $\Phi(x,y,0)$ that undergo the interaction we are now considering, the available initial positions $\{X(0),Y (0)\}$ are initially distributed in such a way that the pointer $Y(t_f )$ shows an apparent correlation with $X(t_f )$ in spite of both variables $x$ and $y$ evolving independently. For example, imagine that our special universe only contains the following 5 initial pairs of particles positions that are deterministically associated with the final pairs of particle positions indicated:

\begin{figure}
\includegraphics[width=0.824\columnwidth]{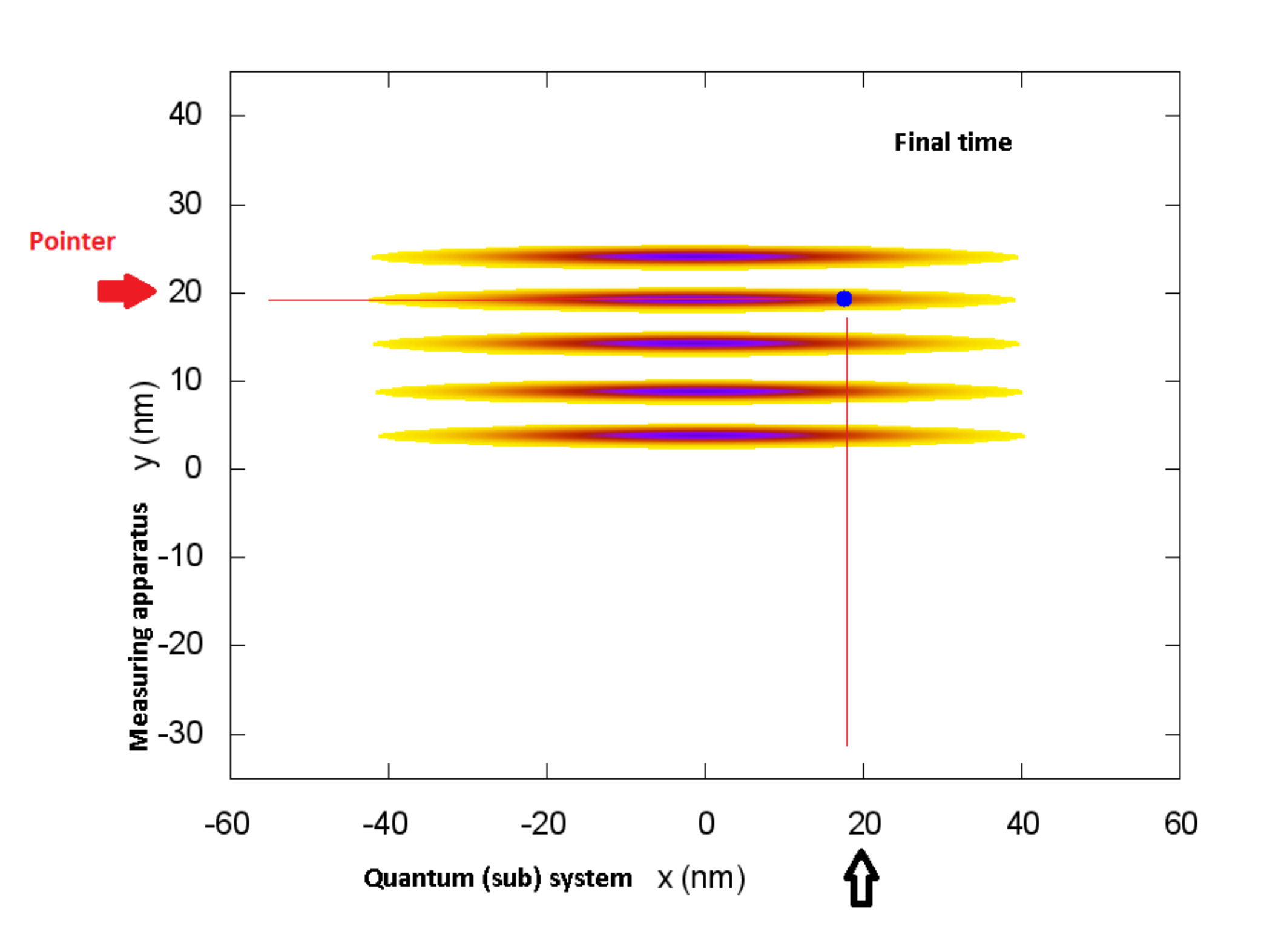}
\caption{Probability density of the (effective) wave function  $\Phi(x,y,t_f)$ at the final time $t_f$ and (filled-in circle) the corresponding (Bohmian) position $\{X(t_f),Y(t_f)) \}$ in the two-dimensional configuration space for one particular experiment. The evolution of   $\Phi(x,y,t_f)$ implies a perturbation in the $y$ direction, but an unperturbed evolution in the $x$ direction.  Such an  interaction cannot be considered as a good measurement of the position because, due to equivariance and the guidance law of effective wave functions, eq. (\ref{velo_eff}), the final value of the pointer $Y(t_f)=20 \; nm$ is compatible with many different positions of the system $X(t_f)$.  (Color figure online).}
\label{fig3}
\end{figure}

\begin{eqnarray}
\{X(0)&=&2\;nm,Y(0)=-27\;nm\} \rightarrow \{X(t_f)=2\;nm,Y(t_f)=2\;nm\}\nonumber\\
\{X(0)&=&8\;nm,Y(0)=-26\;nm\} \rightarrow \{X(t_f)=8\;nm,Y(t_f)=8\;nm\}\nonumber\\
\{X(0)&=&14\;nm,Y(0)=-25\;nm\} \rightarrow \{X(t_f)=14\;nm,Y(t_f)=14\;nm\}\nonumber\\
\{X(0)&=&20\;nm,Y(0)=-24\;nm\} \rightarrow \{X(t_f)=20\;nm,Y(t_f)=20\;nm\}\nonumber\\
\{X(0)&=&26\;nm,Y(0)=-23\;nm\} \rightarrow \{X(t_f)=26\;nm,Y(t_f)=26\;nm\}\nonumber
\end{eqnarray}
In fact, the apparatus is not doing anything; the initial conditions of our universe are so special that it seems that there is a perfect correlation between $X(t_f)$ and $Y(t_f)$: when the system is in $X(0)=2\;nm$ the final pointer indicates $Y(t_f)=2\;nm$ without distorting the wave function of the system, when the system is in $X(0)=8\;nm$ the final pointer indicates $Y(t_f)=8\;nm$ without distorting the system wave-function, and so on. Such a universe is certainly possible, but highly atypical. Assuming that such an atypical universe (or other universes with similar pathologies) can be disregarded, we have shown that a measurement of the initial position that does not disturb the wave function is not possible.

Summing up, the results in subsections \ref{sec221}, \ref{sec222} and \ref{sec223} show the fundamental differences between classical and Bohmian mechanics regarding the possibility of experimentally ascertaining the initial conditions of a system. Both are deterministic theories but, while according to the former we can in principle measure the initial conditions and possibly make a precise prediction of the future evolution of a system, this is not the case according to the latter. 

      We can now see these same results under a new light. Since in Bohmian mechanics we cannot measure the state of a system, this entails that we do not have any means to experimentally select or produce a set of systems that are guaranteed to be identical---in the same Bohmian state. Imagine that we have a collection of electrons all prepared so that their effective wave function is the same. From the standpoint of the orthodox quantum-mechanical approach (and of any quantum theory that assumes that the wave function alone provides a \emph{complete} description of a system), these electrons are identical as they have the same wave function. If we measure a property and we obtain different results, it will be puzzling. We are then forced to accept indeterminism and that there is no cause of the different results. From the standpoint of Bohmian mechanics, however, the electrons are not identical: even if they have the same effective wave function, their positions (relative to the wave function) need not be the same. Therefore, there is no puzzle if they behave differently and produce different results upon measurement. Different initial (Bohmian) states give different final results! 
      
      Given that we cannot make a finer-grained selection procedure (discerning at the same time the wave function and the positions), the fact that we cannot make precise predictions concerning the future evolution of each electron is naturally accounted for: we cannot do so because we lack information on the complete state of each individual system. As we will see in the next section, the most we can do is adopt a statistical treatment of our ensemble, making a guess as to the distribution of the positions given the wave function.

\
\subsubsection{Measurement of the wave function from weak values}
\label{sec224}

On the one hand, we have explained in subsection \ref{sec222} that the wave function cannot be measured. On the other hand, we have shown in subsection \ref{sec223} that, even when the wave function is properly ``prepared'',  a measurement of the position of a particle necessarily perturbs the ``prepared'' wave function (unless the latter is very close to a delta function in the position representation). These results concerning our ignorance of the initial state of the wave function and the initial positions of the particles imply an unavoidable unpredictability in the deterministic Bohmian theory. At this point, the reader may wonder whether these conclusions are in conflict with the ``weak measurement" techniques that have recently attracted a lot of attention both from a theoretical and an experimental point of view \cite{weak1,weak2,weak3}.

In particular, the work entitled ``Direct measurement of the wave function'' \cite{weak2} shows the first experimental reconstruction of the transverse spatial wave function of a photon. The experimental procedure for such a reconstruction is the following. Two consecutive measurements are done in a quantum system with a well prepared wave function. The first measurement performs a \emph{weak} measurement of the position, which implies a \emph{weak} perturbation of the wave function. The second measurement is done on the previously weakly perturbed wave function through a projective---strong---measurement of the momentum that finally \emph{collapses} the wave function. The above two consecutive measurements are repeated in many experiments, each one with an initial wave function ``prepared" in an identical way. Then, from all the experiments, only the experiments whose measured momentum gives a particular value are selected (and the rest of experiments are disregarded). The average value of the measured position computed only from the set of experiments that have been previously selected allows to reconstruct the wave function.  A discussion of this experimental procedure can be found in  \cite{weak2,norsen2}. The average value obtained from the above procedure (that provides information beyond that obtained from standard projective---strong---measurements) is called the ``weak value''.

Does the so-called ``Direct measurement'' of the wave function in Ref. \cite{weak2} contradict the main result of subsection \ref{sec222}? Obviously, no.  Throughout this paper (unless otherwise indicated), we refer to a measurement as the procedure to get an outcome from a single experiment. The value of a measurement of a single experiment can be related to a position of a (measuring apparatus) pointer while, by its own construction, a weak value cannot be related with the position of a pointer in a unique experiment. Therefore, the result of Ref. \cite{weak2} is not the type of ``direct measurement'' that we were dealing with in this paper. The authors of Ref. \cite{weak2} do also use the word ``direct'' in their title to differentiate their experiment with weak values from another experimental technique, quantum tomography\cite{tomography}, which estimates the wave function from a collection of projective (strong) measurements of observables. Each observable in this tomographic technique is obtained from a different experiment with an identically prepared wave function. We prefer here to use the words ``direct measurement'' when the information is obtained from an individual quantum system in a unique experiment (with a single- or multi-time measurement). With our definition, extracting information of the wave function from quantum tomography\cite{tomography} or from weak values\cite{weak1,weak2,weak3} cannot be considered a direct measurement of the wave function, because it is not done in a single experiment. For the same reason, experiments involving weak values of the (Bohmian) velocity cannot be considered as direct measurements of the (Bohmian) trajectories \cite{photons,fabio,traj2}.

\section{A statistical discussion of randomness}
\label{sec3}

When it comes to discussing phenomena such as uncertainty, unpredictability and randomness from the standpoint of Bohmian mechanics, most authors start with the quantum equilibrium hypothesis which is, in fact, the essential statistical consideration assumed in Bohmian mechanics in order to determine the initial positions of particles probabilistically. As we will see, this hypothesis is crucial for establishing the empirical content of the theory and it guarantees the empirical equivalence of Bohmian mechanics with ordinary non-relativistic quantum mechanics. Yet here, we have preferred a different approach, showing first that certain limits on what can be measured arise in Bohmian mechanics. This has been done without explicitly invoking the details of the quantum equilibrium hypothesis and focusing instead on the dynamical laws of the theory. These limits---which do not arise in classical mechanics---in turn imply limits on what can be predicted. 

We ended the last section with the idea that if we have a collection of systems prepared similarly (with the same effective wave function), then we cannot control their positions, since measuring the positions would perturb the wave function. While it could therefore seem as if no prediction can actually be made, the quantum equilibrium hypothesis comes to our rescue with an explicit suggestion of how the positions of systems with the same wave function are statistically distributed. In what follows, when describing the quantum equilibrium hypothesis and its implications, we are reviewing in general the work of D\"urr, Goldstein and Zangh\`i from 1992 in Ref. \cite{Durr1992}

\subsection{The quantum equilibrium hypothesis}
\label{sec31}

The quantum equilibrium hypothesis can be stated as follows:
\begin{quote}
\textbf{Quantum equilibrium hypothesis:} For an ensemble of identical systems, each having the wave function $\phi$, the empirical distribution of the configuration of the particles  $\rho$ is given by $\rho=|\phi|^2$. 
\end{quote}

All quantum experiments (without exception) provide strong empirical support for this hypothesis. It is well known that for any quantum theory to be compatible with the empirical results of all available data regarding quantum phenomena, the Bohmian theory among them, it has to satisfy the Born rule. This rule states that, if a system has a well-known (effective) wave function $\phi(x,t)$  at time $t$, the probability of finding particles with configuration $x$ in volume $dx$ is equal to $\rho(x,t)dx=|\phi(x,t)|^2dx$.  In consequence, if we want to ensure that quantum mechanics and Bohmian mechanics have the same empirical content, the Born rule has to hold in Bohmian mechanics as well. But in Bohmian mechanics, the measured configuration always corresponds with the actual configuration of the system. Therefore, the empirical equivalence of Bohmian mechanics and quantum mechanics is satisfied when the quantum equilibrium hypothesis is considered.

\subsection{The justification of the quantum equilibrium hypothesis}
\label{sec32}

For Bohmian mechanics to predict the results determined experimentally, we have to assume the quantum equilibrium hypothesis, which enters the theory either as a postulate or as a consequence of some deeper physical consideration. An insightful justification for the quantum equilibrium hypothesis has been proposed by D\"urr et al. \cite{durr13,Durr1992}. They argue that the quantum equilibrium hypothesis is just a consequence of living in a \emph{typical} universe. Given that Eqs. (\ref{scho}) and (\ref{velo}) are deterministic, picking an initial configuration for all the Bohmian particles, $\mathbf{X}(0) = \{ X_1(0),X_2(0),...,X_N(0)\}$, amounts to picking one complete possible history of the universe. Consider a history with many subsystems of the universe that, at different places and times, have the same conditional wave function, $\psi$ (with respect to each one's own subsystem coordinates). We will say that the quantum equilibrium hypothesis is satisfied in this particular history if the actual empirical distribution of the configurations of these subsystems suitably approximates the distribution $\rho=|\psi|^2$. Using the law of large numbers and assuming that initial configurations (and therefore histories) are weighted with the measure given by $\rho(\mathbf{x},0)=|\Psi(\mathbf{x},0)|^2$, D\"urr et al. show that \emph{most} initial configurations lead to histories that satisfy the quantum equilibrium hypothesis (if we understand  $\rho(\mathbf{x},0)=|\Psi(\mathbf{x},0)|^2$ as giving a measure of typicality over the initial global configuration of particles; that is, over the set of Bohmian histories)\footnote{The interpretation of the measure $\rho(x,0) = |\Psi(x, 0)|^2$ is a subtle and complicated matter. Understood literally as a probabilistic distribution, it invites one either to think of a supernatural being playing with the initial conditions or to a subjectivist reading. We do not favor either of these. With respect to this question, Bell eloquently asserts: ``A single configuration of the world will show statistical distributions over its different parts. Suppose, for example, this world contains an actual ensemble of similar experimental set-ups. [...] It follows from the theory that the `typical' world will approximately realize quantum mechanical distributions over such approximately independent components. The role of the hypothetical ensemble is precisely to permit definition of the word `typical' .'' (\cite{bell}, p. 129). Here we want to go along with Bell in considering that $\rho(x, 0)$ is a measure of typicality---not a probabilistic measure---and to stress that the only role of the hypothetical ensemble is precisely to permit definition of the word `typical.' For more details on this issue, see \cite{Durr1992}.}. Note that this account is not circular, since D\"urr et al. derive the quantum equilibrium hypothesis (which is a constraint applied to subsystems of the universe with well-defined effective or conditional wave functions) from the assumption that the initial configuration of the whole universe is distributed according to $\rho(\mathbf{x},0)=|\Psi(\mathbf{x},0)|^2$, which is a constraint on Bohmian histories\footnote{A more detailed discussion of this argument can be found in this Special Issue in the article of Michael Esfeld and co-authors.}.

\subsection{Absolute uncertainty}
\label{sec33}

If it is assumed that the configuration of the whole universe at the initial time is distributed according to $\rho(\mathbf{x},0)=|\Psi(\mathbf{x},0)|^2$ and given that $|\Psi(\mathbf{x},t)|^2$ is equivariant, the following \textbf{fundamental conditional probability formula} can easily be derived:

\begin{equation}
\text{P}(X(t) \in \{x,x+dx\} | \mathbf{y}=\mathbf{Y}(t)) = \frac{|\Psi(x,\mathbf{Y}(t),t)|^2dx}{\int |\Psi(x,\mathbf{Y}(t),t)|^2 dx } = |\psi(x,t)|^2 dx.
\label{cond-prob}
\end{equation}
Recall that here, $x$ is the variable of the (sub)system of interest, while $\mathbf{y}$ represents the variables of the environment and $|\Psi(x,\mathbf{Y}(t),t)|^2$ is the joint distribution of the configuration $\{ X(t) \in \{x,x+dx\} , \mathbf{Y}(t) \}$. In addition, $\psi(x,t)$ is the normalized conditional wave function\footnote{The normalization constant is irrelevant because the Bohmian velocity does not depend on normalization factors.}. 

 It follows from Eq. (\ref{cond-prob}) that even in the case when the exact detailed configuration of the whole environment,  $\mathbf{Y}(t)$, is precisely known  (information which is not accessible via measurements), there is no more information on the position of the subsystem studied than that expressed in the right-hand side of Eq. (\ref{cond-prob}). In other words, even if an experimenter could known exactly all the positions of the particles composing the measuring apparatus or environment associated with a quantum subsystem, the maximum information on the position of the particle that constitutes the subsystem being studied is the modulus squared of its conditional wave function $|\psi(x,t)|^2$. In this sense, D\"urr \emph{et al.} \cite{Durr1992} have called the uncertainty implied by Eq. (\ref{cond-prob}) \emph{absolute uncertainty}. All the empirical statistical contents of Bohmian mechanics follow from this formula. 

It is worthwhile making the following point: the absolute uncertainty and conditional probability formula just introduced do not imply that precise information on the configuration of the subsystem $X(t)$ cannot be obtained. The crucial aspect expressed in Eq. (\ref{cond-prob}) is that such knowledge (precise information on the subsystem $X(t)$) must be mediated by $\psi(x,t)$. But in Bohmian mechanics, $\psi(x,t)$ does not merely represents our knowledge of the system; it also has an important and crucial dynamical aspect, i.e. it guides the motion of the particles. This implies that the absolute uncertainty just introduced above embodies, when the specific dynamics is considered, \emph{absolute unpredictability}. 

In section \ref{sec2}, we use dynamical arguments to show that we cannot know the position and the wave function of a subsystem of interest at the same time, because measuring the position always disturbs the wave function. In what follows, we want to show that this is in perfect agreement with the statistical considerations introduced above in this subsection. In order to see this, we again consider the subsystems $x$ and $y$ mentioned in the numerical example of subsection \ref{sec223}, with an initial (conditional or effective) wave function $\phi(x,0)=\Phi(x,Y(0),0)$  with some spread along the $x$-axis, as represented in Fig. \ref{fig1}. We will also consider that the interaction between the subsystems $x$ and $y$ constitutes a measurement of the position of $x$: after the interaction, the final position of the pointer, $Y(t_f)$, is suitably correlated with $X(t_f)$, so we can infer the latter with precision by looking at the former.  Finally, let us assume (for the sake of the argument) that the (effective) wave function of the system at the final time is equal to that function at the initial time, i.e. $\phi(x,t_f)=\phi(x,0)$. Now, this last condition contradicts Eq. (\ref{cond-prob}) for the following reason. Given the knowledge of the environment, we know the position of $x$ perfectly. For instance, the probability distribution of the position of $x$ (conditioned by our knowledge that the final pointer position is $Y(t_f) \approx 20$ nm) is approximately $\delta(x-20 \; nm)$. From Eq. (\ref{cond-prob}), it then follows that the (conditional) wave function after the measurement is, indeed, a delta function. However, above we also require that $\phi(x,t_f)=\phi(x,0)$. The two requirements are incompatible, because $\delta(x-20 \; nm) \neq |\phi(x,0)|^2$ (unless our initial wave function was in fact a delta function, which contradicts our supposition). 

As can be seen from this example, it follows from the fundamental conditional probability formula that if, after an experiment, by knowing the position of the pointer we can infer with precision the position of the object, then the conditional wave function of the object after the experiment must approximate a delta function regardless of what the initial conditional wave function was.

\section{Randomness, measurement and equilibrium}
\label{sec4}

In section \ref{sec2}, we relate quantum randomness with our inability to measure the initial conditions of a quantum system. Next, in section \ref{sec3} we claimed that quantum randomness arises because our universe is in quantum equilibrium. These two clarifications of the concept of quantum randomness may seem unconnected. In this section, we want to briefly argue that this is not the case and that the notion of measurement is indeed closely related with the statistical notions of equilibrium and non-equilibrium. 

\subsection{Measurement and equilibrium}
\label{sec41}

Measuring implies obtaining knowledge of a system through the use of an apparatus. Independently of whether we are considering a classical or a quantum measurement, such knowledge requires the existence of a correlation between the system and the apparatus; a correlation that must be stable and robust enough to lead to the formation of a permanent record (e.g., a black spot on a photographic plate, a computer printout, etc.). We usually take the possibility of knowledge (that is, the possibility of the existence of strong and robust correlations between different subsystems) for granted. Yet the fact that this type of correlations does exist depends on how our universe works; and its working could easily have been otherwise.

Consider, for instance, the toy-model of a classical measurement that we introduced in subsection \ref{sec221}. The very fact that there is a macroscopic pointer that correlates with some property,  $A (x, p_x)$, of a system, as represented by the Hamiltonian in Eq. (\ref{al1}), presupposes a scenario that is clearly out of equilibrium. If the whole universe were in thermodynamic equilibrium, no such macroscopic correlations could arise. Therefore, the very possibility of measuring is related to the equilibrium conditions that hold in a particular situation.

When it comes to thermodynamic equilibrium, the universe can be considered to be a sea of non-equilibrium with some islands of equilibrium. This fact regarding our universe allows for the ubiquitous existence of measuring-type interactions, such as those described by the Hamiltonian in Eq. (\ref{al1}). When it comes to quantum equilibrium, however, the situation is not the same. All the empirical evidence we have suggests that our universe is \textit{globally} in quantum equilibrium. Therefore, \textit{all} systems are in quantum equilibrium and the limitations that this condition imposes upon measurements admit no exception. In the next subsection, we explore these limitations again.

\subsection{Quantum equilibrium and position measurement}
\label{sec42}

We have shown that in classical mechanics it is possible to establish a perfect correlation between a pointer and the position of a particle. Here, we now argue why the same measurement of position is not possible in a universe in quantum equilibrium. 

A quantum measurement of position also requires a correlation between $X(t_f)$ and $Y(t_f)$. The two positions are, however, in quantum equilibrium. Because of this equilibrium, when we predict the (Bohmian) position of the particle we have to treat  $X(t_0)$ and $Y(t_0)$ as random variables distributed according to $|\Phi(x,y,t_0)|^2$. Therefore, different experiments are associated with different random initial positions that, via the guiding law in equation (\ref{velo_eff}), will correspond to different $X(t_f)$ and $Y(t_f)$ at the final time $t_f$. By equivariance, after the experiment, the probability distribution of $X(t_f)$ and $Y(t_f)$ is given by $|\Phi(x,y,t_f)|^2$. Therefore, the only way to really establish a close correlation between  $X(t_f)$ and $Y(t_f)$ in all experiments is through the quasi-delta function shown in Fig. \ref{fig2}. In contrast, quantum equilibrium implies that the type of wave function depicted in Fig. \ref{fig3} will never provide perfect correlation between the system $X(t)$ and the pointer $Y(t)$ (the same value of the pointer indicating $Y(t_f)$ is compatible with many different positions of the system in different experiments). 

Finally, let us stress again that, as opposed to other types of equilibrium (thermal, electrostatic or thermodynamical), quantum equilibrium does not require \lq\lq{}relaxation times\rq\rq{}. As evidenced by all experiments, our universe is in quantum equilibrium, and all subsystems satisfy the quantum equilibrium hypothesis at any time. This is just a consequence of the equivariant property of the universal wave function discussed in section \ref{sec2}. Thus, there are unavoidable limitations to our knowledge of the positions of particles. In fact, once the wave function is prepared, there is an \emph{absolute uncertainty} regarding the positions of the particles. 

\section{Conclusions}
\label{sec5}

We have tried to answer the question: \textit{ How does quantum uncertainty emerge from deterministic Bohmian mechanics?}  The equations of motion of classical and Bohmian mechanics are both fully deterministic. However, to be able to determine the output of a (classical or quantum) experiment with certainty, we have to know the initial conditions of the system. We have seen that in classical mechanics, there are many scenarios where the initial conditions can be measured and therefore we can predict the future evolution of the system with certainty. Meanwhile, in other classical systems (such as chaotic systems or those involving a very large number of particles), we cannot determine the initial conditions with enough precision to make predictions about the future with certainty. In Bohmian mechanics, the initial state of a system (including both the positions of the particles and the wave function) cannot be determined experimentally. Therefore, as our knowledge of the initial conditions is constrained, we cannot predict the future evolution of the quantum system with certainty. 

The reader may realize that the foregoing answer as to how a (classical or quantum) deterministic theory becomes unpredictable is quite trivial. However, a look at the history of science shows that the unavoidable randomness of quantum systems opened an intense debate on the impossibility of using explanations based on deterministic laws for quantum phenomena.  The famous (and incorrect) Von Neumann  theorem\cite{neuman} and the related \lq\lq{}impossibility proofs\rq\rq{} are the most evident examples of the intensity of the arguments against deterministic quantum theories\cite{bell}. In section \ref{sec2} we show that the Bohmian dynamical laws do not allow the measurement of a wave function in a single experiment. Even assuming a \textit{preparation} of the wave function (not a measurement), then the same laws do not allow us to determine the position of the particle without modifying the wave function (unless the initial wave function is a position eigenstate). Without the possibility of having experimental access to the initial conditions, the deterministic Bohmian theory becomes a theory that involves uncertainty in its predictions.  

We have not only provided an answer to the question regarding how the unpredictability of the results of quantum experiments can follow from a  deterministic theory, but we have also introduced the fundamental statistical hypothesis that quantifies the amount of randomness that appears in quantum experiments. In classical systems, when the initial conditions are inaccessible, classical statistical mechanics provides probabilistic information on typical initial conditions. In quantum systems, since the initial positions of particles are always inaccessible via measurement (without converting the initial wave function into a position eigenstate), the quantum equilibrium hypothesis determines the probability of different initial positions. This hypothesis is merely a consequence of the fact that our universe is always in quantum equilibrium (in the sense discussed in section \ref{sec3}).

All the dynamical and statistical insights of Bohmian mechanics can be summarized in the fundamental conditional probability formula, which states that the maximum information on the position of a particle can be obtained from the modulus squared of its (conditional) wave function: $|\psi(x,t)|^2$. Stated simply, any knowledge of the particle positions must unavoidably be mediated by the wave function. This imposes an \textit{absolute uncertainty} on quantum mechanics, which is the fundamental key to understanding how  Bohmian mechanics, despite being deterministic, can  account for all quantum predictions, including quantum randomness and uncertainty.

\section*{Acknowledgments}

This work has been partially supported by the Fondo Europeo de Desarrollo Regional (FEDER) and Ministerio de Econom\'{i}a y Competitividad through the Spanish Projects No. TEC2012-31330 and No. TEC2015-67462-C2-1-R, the Generalitat de Catalunya (2014 SGR-384), and by the European Union Seventh Framework Program under the Grant Agreement No. 604391 of the Flagship initiative “Graphene-Based Revolutions in ICT and Beyond”.  A.S.'s work was supported by the project FFI2012-37354 funded by the Spanish Ministry of Economy and Competitiveness. N.Z. was supported in part by the Italian \textit{Istituto Nazionale di Fisica Nucleare}.

\end{document}